\documentstyle[editedvolume]{crckapb} 

\newcommand{\qq}{\qquad}

\newcommand{\msol}{\ifmmode {M_\odot}\else {$M_\odot$}\fi}
\newcommand{\lsol}{\ifmmode {L_\odot}\else {$L_\odot$}\fi}
\newcommand{\etal}{{\it et al.\ }}
\newcommand{\psqcm}{\ifmmode {\>{\rm cm}^{-2}}\else {cm$^{-2}$}\fi}
\newcommand{\pcubcm}{\ifmmode {\>{\rm cm}^{-3}}\else {cm$^{-3}$}\fi}
\newcommand{\xieff}{$\xi_{\rm eff}$}

\newcommand{\kms}{\ifmmode {\>{\rm km\ s}^{-1}}\else {km s$^{-1}$}\fi}
\newcommand{\iintensity}{\ifmmode{{\rm erg\ cm}^{-2}{\rm\ s}^{-1} {\rm\
sr}^{-1}}\else{erg cm$^{-2}$ s$^{-1}$ sr$^{-1}$}\fi}
\def\gtrapprox{\;\lower 0.5ex\hbox{$\buildrel >
    \over \sim\ $}}             
\def\lessapprox{\;\lower 0.5ex\hbox{$\buildrel < \over \sim\ $}}
\newcommand\epuv{\ifmmode{{\rm erg\ cm}^{-3}}\else {erg
cm$^{-3}$}\fi}
\newcommand\einpuv{\ifmmode{{\rm erg\ cm}^{-3}{\rm\ s}^{-1}}\else {erg
cm$^{-3}$ s$^{-1}$}\fi}
\newcommand\lum{\ifmmode{{\rm erg\ s}^{-1}}\else {erg s$^{-1}$}\fi}
\def\etal{{\it et al.\ }}

\def\eg{{\it e.g.,\ }}
\def\cf{{\it cf.\ }}
\def\ie{{\it i.e.,\ }}
\def\mslpyr{M_\odot{\rm\ yr}^{-1}}
\def\cmsq{\ifmmode {\>{\rm\ cm}^2}\else {cm$^2$}\fi}
\def\psqcm{\ifmmode {\>{\rm cm}^{-2}}\else {cm$^{-2}$}\fi}
\def\psqpc{\ifmmode {\>{\rm pc}^{-2}}\else {pc$^{-2}$}\fi}
\def\pcsq{\ifmmode {\>{\rm\ pc}^2}\else {pc$^2$}\fi}
\def\intensity{\ifmmode{{\rm erg\ cm}^{-2}{\rm\ s}^{-1}
      {\rm\ Hz}^{-1}{\rm\ sr}^{-1}}
      \else {erg cm$^{-2}$ s$^{-1}$ Hz$^{-1}$ sr$^{-1}$}\fi}
\def\iintensity{\ifmmode{{\rm erg\ cm}^{-2}{\rm\ s}^{-1} {\rm\
sr}^{-1}}\else
      {erg cm$^{-2}$ s$^{-1}$ sr$^{-1}$}\fi}
\def\flux{\ifmmode{{\rm erg\ cm}^{-2}{\rm\ s}^{-1}}\else {erg
cm$^{-2}$ s$^{-1}$}\fi}
\def\fluxdensity{\ifmmode{{\rm erg\ cm^{-2}\ s^{-1}\ Hz^{-1}}}\else {erg
cm$^{-2}$ s$^{-1}$ Hz$^{-1}$}\fi}
\def\lum{\ifmmode{{\rm erg\ s}^{-1}}\else {erg s$^{-1}$}\fi}
\def\Lbol{\ifmmode {L_{\rm bol}}\else {$L_{\rm bol}$}\fi}
\def\Lacc{\ifmmode {L_{\rm acc}}\else {$L_{\rm acc}$}\fi}
\def\Lmech{\ifmmode {L_{\rm mech}}\else {$L_{\rm mech}$}\fi}
\def\Ledd{\ifmmode {L_{\scriptscriptstyle {\rm EDD}}}\else 
{$L_{\scriptscriptstyle {\rm EDD}}$}\fi}
\def\Lism{\ifmmode {L_{\scriptscriptstyle {\rm ISM}}}\else 
{$L_{\scriptscriptstyle {\rm ISM}}$}\fi}
\def\Nlyc{\ifmmode {N_{\scriptscriptstyle {\rm LyC}}}\else 
{$N_{\scriptscriptstyle {\rm LyC}}$}\fi}
\def\nusn{\ifmmode {\nu_{\scriptscriptstyle {\rm SN}}}\else 
{$\nu_{\scriptscriptstyle {\rm SN}}$}\fi}
\def\nuco{\ifmmode {\nu_{\scriptscriptstyle {\rm CO}}}\else 
{$\nu_{\scriptscriptstyle {\rm CO}}$}\fi}
\def\nuhcn{\ifmmode {\nu_{\scriptscriptstyle {\rm HCN}}}\else 
{$\nu_{\scriptscriptstyle {\rm HCN}}$}\fi}
\def\muco{\ifmmode {\mu_{\scriptscriptstyle {\rm CO}}}\else 
{$\nu_{\scriptscriptstyle {\rm CO}}$}\fi}
\def\muhcn{\ifmmode {\mu_{\scriptscriptstyle {\rm HCN}}}\else 
{$\nu_{\scriptscriptstyle {\rm HCN}}$}\fi}
\def\Lsn{\ifmmode {L_{\scriptscriptstyle {\rm SN}}}\else 
{$L_{\scriptscriptstyle {\rm SN}}$}\fi}
\def\Esn{\ifmmode {E_{\scriptscriptstyle {\rm SN}}}\else 
{$E_{\scriptscriptstyle {\rm SN}}$}\fi}
\def\Eob{\ifmmode {E_{\scriptscriptstyle {\rm OB}}}\else 
{$E_{\scriptscriptstyle {\rm OB}}$}\fi}
\def\Edotob{\ifmmode {\dot E_{\scriptscriptstyle {\rm OB}}}\else 
{$\dot E_{\scriptscriptstyle {\rm OB}}$}\fi}
\def\Eagn{\ifmmode {E_{\scriptscriptstyle {\rm AGN}}}\else 
{$E_{\scriptscriptstyle {\rm AGN}}$}\fi}
\def\Exagn{\ifmmode {E_{\scriptscriptstyle {\rm AGN}}^x}\else 
{$E_{\scriptscriptstyle {\rm AGN}}^x$}\fi}
\def\Eb{\ifmmode {E_{\scriptscriptstyle {\rm B}}}\else 
{$E_{\scriptscriptstyle {\rm B}}$}\fi}
\def\mdotsn{\ifmmode {\dot M_{\scriptscriptstyle {\rm SN}}}\else 
{$\dot M_{\scriptscriptstyle {\rm SN}}$}\fi}
\def\Mbh{\ifmmode {M_{\scriptscriptstyle {\rm BH}}}\else 
{$M_{\scriptscriptstyle {\rm BH}}$}\fi}
\def\Pism{\ifmmode {P_{\scriptscriptstyle {\rm ISM}}}\else 
{$P_{\scriptscriptstyle {\rm ISM}}$}\fi}
\def\tPism{\ifmmode {\tilde P_{\scriptscriptstyle {\rm ISM}}}\else 
{$\tilde P_{\scriptscriptstyle {\rm ISM}}$}\fi}
\def\tPturb{\ifmmode {\tilde P_{\rm turb}}\else 
{$\tilde P_{\rm turb}$}\fi}
\def\nh{\ifmmode {n_{\scriptscriptstyle {\rm H}}}\else 
{$n_{\scriptscriptstyle {\rm H}}$}\fi}
\def\nh{\ifmmode {n_{\scriptscriptstyle {\rm H}}}\else 
{$n_{\scriptscriptstyle {\rm H}}$}\fi}
\def\nhtwo{\ifmmode {n_{\scriptscriptstyle {\rm H_2}}}\else 
{$n_{\scriptscriptstyle {\rm H_2}}$}\fi}
\def\xhtwo{\ifmmode {x_{\scriptscriptstyle {\rm H_2}}}\else 
{$x_{\scriptscriptstyle {\rm H_2}}$}\fi}
\def\xco{\ifmmode {x_{\scriptscriptstyle {\rm CO}}}\else 
{$x_{\scriptscriptstyle {\rm CO}}$}\fi}
\def\xhcn{\ifmmode {x_{\scriptscriptstyle {\rm HCN}}}\else 
{$x_{\scriptscriptstyle {\rm HCN}}$}\fi}
\def\xc{\ifmmode {x_{\scriptscriptstyle {\rm C}}}\else 
{$x_{\scriptscriptstyle {\rm C}}$}\fi}
\def\xcp{\ifmmode {x_{\scriptscriptstyle {\rm C^+}}}\else 
{$x_{\scriptscriptstyle {\rm C^+}}$}\fi}
\def\Nh{\ifmmode {N_{\scriptscriptstyle {\rm H}}}\else 
{$N_{\scriptscriptstyle {\rm H}}$}\fi}
\def\Nc{\ifmmode {N_{\scriptscriptstyle {\rm C}}}\else 
{$N_{\scriptscriptstyle {\rm C}}$}\fi}
\def\Natt{\ifmmode {N_{\rm att}}\else {$N_{\rm att}$}\fi}
\def\mh{\ifmmode {m_{\scriptscriptstyle {\rm H}}}\else 
{$m_{\scriptscriptstyle {\rm H}}$}\fi}
\def\Edotkin{\ifmmode {\dot E_{\rm kin}}\else {$\dot E_{\rm kin}$}\fi}
\def\Ekin{\ifmmode {E_{\rm kin}}\else {$E_{\rm kin}$}\fi}
\def\tdiss{\ifmmode {t_{\rm diss}}\else {$t_{\rm diss}$}\fi}
\def\pyr{\ifmmode {\>{\rm\ yr}^{-1}}\else {yr$^{-1}$}\fi}
\def\psec{\ifmmode {\>{\rm\ s}^{-1}}\else {s$^{-1}$}\fi}
\def\epstar{\ifmmode {\epsilon_*}\else {$\epsilon_*$}\fi}
\def\frem{\ifmmode {f_{\rm rem}}\else {$f_{\rm rem}$}\fi}
\def\fdis{\ifmmode {f_{\rm dis}}\else {$f_{\rm dis}$}\fi}
\def\nt{\ifmmode{{\rm cm^{-3}\ K}}\else {cm$^{-3}$ K}\fi}
\def\xieff{\ifmmode {\xi_{\rm eff}}\else {$\xi_{\rm eff}$}\fi}
\def\jonezero{$J=1\rightarrow 0$}
\def\fpah0{\ifmmode {f_{\scriptscriptstyle {\rm PAH^o}}}\else 
{$f_{\scriptscriptstyle {\rm PAH^o}}$}\fi}
\def\fcp{\ifmmode {f_{\scriptscriptstyle {\rm C^+}}}\else 
{$f_{\scriptscriptstyle {\rm C^+}}$}\fi}
\def\be{\begin{equation}}
\def\ee{\end{equation}}
\def\bea{\begin{eqnarray}}
\def\eea{\end{eqnarray}}
\def\plotfiddle#1#2#3#4#5#6#7{\centering \leavevmode
\vbox to#2{\rule{0pt}{#2}}
\includegraphics{#1}}

\begin{opening}
\title{THE IMPACT OF STAR FORMATION AND ACTIVE NUCLEI
ON THE INTERSTELLAR MEDIUM IN ULTRALUMINOUS INFRARED GALAXIES}

\author{PHILIP R. MALONEY}
\institute{Center for Astrophysics and Space Astronomy\\
           University of Colorado, Boulder, CO 80309-0389}
\end{opening}

\runningtitle{THE ISM IN ULIRGS}

\begin{document}

\begin{abstract}
The energy input into the interstellar medium in Ultraluminous
Infrared Galaxies (ULIRGs) is enormous, regardless of the nature of
the power source. I discuss some of the major consequences for the
structure and energetics of the ISM in these galaxies. 
Observationally, the column densities in the nuclear regions of ULIRGs
are known to be very high, which makes distinguishing starbursts from
AGN quite difficult. The level of energy and momentum injection means
that the pressure in the ISM must be extremely high, at least $3-4$
orders of magnitude larger than in the local ISM or typical giant
molecular clouds. It also means that the luminosity of GMCs in ULIRGs
must be very high, as they must radiate many times their binding
energy over their lifetimes. I briefly review the influence which
X-ray irradiation can have on the ISM in AGN-powered ULIRGs. Finally,
I show that the presence of PAH features in ULIRGs does not imply that
they must be starburst-dominated, since at the column densities and
pressures typical of the ISM in ULIRGs PAHs can survive even at tens
of parsec distances from the AGN.
\end{abstract}

\section{Introduction}
The observed characteristics of ULIRGs (luminosities $L_{\rm IR}\ge
10^{12}\lsol$, size scales $r\lessapprox\;$a few hundred pc) imply typical
energy densities $u\sim 10^{-8}$ erg cm$^{-3}$. This is 3--4 orders of
magnitude larger than the typical energy densities (in gas, photons,
and cosmic rays) in the interstellar medium (ISM) in the solar
neighborhood (\eg Jura 1987). In this review, I will discuss how both
the production and release of such energies is expected to impact the
ISM. I will concentrate on those aspects most relevant to ULIRGs, \ie
very powerful starbursts and luminous active galactic nuclei
(AGN). Due to space limitations, the discussion is necessarily
terse. In particular, the referencing is by no means complete, and
should be regarded largely as a pointer to the relevant literature (a
good place to start is the recent review by Sanders \& Mirabel 1996);
my apologies to those whose contributions have of necessity been left
out. 

\section{Basic Parameters}
\subsection{Starburst:}
{\it Bolometric Luminosity} The frequency-integrated energy output
from a starburst can be approximated as
\be
\Lbol\sim 10^{10}\left({\dot M\over\mslpyr}\right)\left({\Delta
t\over 10^8\;{\rm yr}}\right)^{2/3}\left({m_l\over
1\;\msol}\right)^\alpha \left({m_u\over 45\;\msol}\right)^{0.37}\;\lsol
\ee
for a constant star formation rate (SFR) burst of age $\Delta t$,
upper and lower mass cutoffs $m_u$ and $m_l$, and $\alpha=0.23$ for
$m_l < 1\;\msol$ (Scoville \& Soifer 1991; see also Smith, Lonsdale,
\& Lonsdale 1998). (Here and frequently throughout this review,
adopted scalings with mass, density, etc. should be regarded with the
caution usually reserved for possibly rabid housepets.) This implies
that powering ULIRGs by star formation requires
\be
\dot M\sim 100\left({\Lbol\over 10^{12}\;\lsol}\right)
\left({\Delta t\over 10^8\;{\rm yr}}\right)^{-0.67} \;\mslpyr
\ee
modulo the details of the assumed initial mass function. Thus ULIRGs
must have SFR in excess of $100\;\mslpyr$, barring extreme assumptions
about the IMF.

\noindent{\it Lyman Continuum Photons} The number of Lyman continuum
($E \ge 13.6$ eV) photons produced per second by a starburst is
approximately (Scoville \& Soifer 1991)
\be
\Nlyc\sim 10^{53}\left({\dot M\over\mslpyr}\right)
\left({m_l\over 1\;\msol}\right)^\alpha \left({m_u\over
45\;\msol}\right)^\beta\psec 
\ee
($\beta=2$, $m_u < 45\;\msol$; $\beta=1.35$, $m_u > 45\;\msol$), where
I have again assumed a constant SFR burst with an age of $10^8$
years. Thus
\be
\Nlyc\sim 10^{55}\left({\Lbol\over 10^{12}\;\lsol}\right)\psec
\ee
for a typical ULIRG. Since the radiation is stellar in origin, the
spectrum of radiation is sharply cutoff above a few tens of eV.

\noindent{\it Supernova Rate} The implied supernova rate for a
starburst-powered ULIRG is
\be
\nusn\sim 1\left({\Lbol\over 10^{12}\;\lsol}\right)\pyr \;.
\ee
This rate is not very sensitive to the assumptions made about the IMF,
since the predecessors of most SNe have $M\sim 10\;\msol$.

\subsection{AGN:}
{\it Bolometric Luminosity} The accretion luminosity for an accreting
black hole (presumed, for many reasons, to be the power source in AGN)
which is converting rest-mass energy to radiation with an
efficiency $\epsilon$ is
\be
\Lacc=1.5\times 10^{12}\left({\epsilon\over 0.1}\right)
\left({\dot M\over \mslpyr}\right)\;\lsol
\ee
independent of the mass of the hole. However, the {\it Eddington
limit} (the luminosity above which radiation pressure will halt
accretion for steady, spherically-symmetric flow) is
\be
\Ledd\simeq 3.4\times 10^{11}\left({\epsilon\over 0.1}\right)
\left({\Mbh\over 10^7\;\msol}\right)\;\lsol .
\ee
If we require $\Lacc < \Ledd$, we must have
\be
\Mbh\gtrapprox 4\times 10^7 \left({\epsilon\over 0.1}\right) 
\left({\dot M\over \mslpyr}\right)\;\msol .
\ee

\noindent{\it Lyman Continuum Photons} The typical spectrum of an AGN
is a power-law in energy, with a flux density
$f_\nu\propto\nu^{-\alpha}$, with $\alpha\sim 1$ (\ie the spectrum is
{\it flat}, with approximately equal energy per decade of photon
energy). For unobscured (Type 1) AGN, the X-ray luminosity $L_x\sim
0.1 \Lbol$, and the spectral index in the hard X-ray regime
$\alpha\approx 0.7$. The fraction of energy emitted in the 2-10 keV
band {\it decreases} slowly with increasing luminosity (see the
summary in Mushotzky, Done, \& Pounds 1993). For a spectrum with $\nu
f_\nu={\rm constant}$ from 1 eV to 100 keV, the number of Lyman
continuum photons is
\be
\Nlyc\sim 10^{55}\left({\Lbol\over 10^{12}\;\lsol}\right)\psec
\ee
essentially identical to the number produced by a starburst. Although
an AGN spectrum extends to far higher energies than that of a
starburst, most of the ionizing photons {\it by number} are between
13.6 and 100 eV, just as in a starburst.

\section{Fueling}
\subsection{Starburst:}
Forming stars at a rate $\dot M_s\;\mslpyr$ requires gas {\it
utilization} at a rate
\be
\dot M={\dot M_s\over \epsilon_*}
\ee
where $\epsilon_*$ is the efficiency of star formation, \ie the
fraction of the mass of a star-forming cloud which is actually
transformed into stars in the cloud's lifetime. The observed value
$\epstar\sim 0.01-0.1$ in Galactic molecular clouds. (Note,
however, that in order to form a bound stellar system, such as a
globular cluster, simple binding energy arguments show that the
efficiency must be much higher, $\epstar\gtrapprox 0.4$; if the
so-called ``proto-globular clusters'' seen, for example, in NGC
4038/4039 [Whitmore \& Schweizer 1995] are in fact bound systems, then
the SFE must have been quite high.) Of this gas used by star
formation, an amount 
\be
\dot M_{\rm eff}={1-\epstar\over\epstar}\fdis\dot M_s+\frem\dot
M_s\equiv f\dot M_s
\ee
is lost from the system. In this equation \fdis\ is the fraction of
the non-stellar remainder of the star-forming cloud which has either
been blown entirely out of the system (\eg in a galactic superwind:
see \S 4), or is either too hot or too dispersed to reform into
star-forming clouds on the characteristic timescale of the burst;
\frem\ is the fraction of the stellar mass that is in the form of
stellar remnants or stars which have masses too low for them to evolve
off the main sequence during the lifetime of the burst. The true gas
usage timescale is thus
\be
\tau_*={M_{\rm gas}\over \dot M_{\rm eff}}= {M_{\rm gas}\over f\dot
M_s}=10^7\;{\rm yr} \left({M_{\rm gas}\over 10^9\;\msol}\right) 
\left({100\;\mslpyr\over \dot M_s}\right) f^{-1}\;.
\ee
The values of both \fdis\ and \frem\ depend strongly on $\Delta
t$. For example, for a burst duration $\Delta t=10^7$ yr, only stars
with $M \gtrapprox 13\;\msol$ have evolved off the main sequence. For
a Miller-Scalo (1979) IMF with $m_l=10$, $m_u=62$, $\frem\sim 0.4$; if
the lower mass limit is reduced to $m_l=5$, $\frem\sim 0.7$. If
$\epstar=0.1$, $\frem=0.5$, and $\fdis=0.5$ (as is suggested by
observations in the case of M82: see \S 4.1), then $f=5$, and gas is
effectively consumed at a much greater rate than that at which it is
formed into stars. It must be emphasized again, however, that the
effective values of \frem\ and \fdis\ are uncertain, depending not
only on the duration of the burst but, for \fdis, on the physical
conditions in the ISM, since this will presumably determine the
recycling time for the gas.

\subsection{AGN:} The fueling requirements for AGN-powered ULIRGs are
comparatively trivial: for the required rate $\dot M \sim 1$
$\mslpyr$, the lifetime of the activity is
\be
\tau_{\scriptscriptstyle{\rm AGN}}\sim 10^9\;{\rm yr}
\left({M_{\rm gas}\over 10^9\;\msol}\right) f_{\rm loss}^{-1}
\ee
where $f_{\rm loss}\ge 1$ takes account of mass that is lost without
accretion in the form of winds, etc. Of course, it is exceedingly
unlikely that the gas reservoir is simply quiescently orbiting the
black hole, waiting to be accreted.

\section{Superbubbles and Superwinds}
\subsection{Starburst:}
Massive stars input both energy and momentum into the ISM, through
radiation, stellar winds and supernova blast waves. In systems with
high rates of star formation, this input can have dramatic impact on
the dynamics of the ISM. In standard stellar wind theory (Weaver \etal
1977), the wind produces a hot shocked bubble in the ambient ISM; the
swept-up ISM at the boundary will collapse to a thin shell provided
the density is high enough for cooling to be efficient.

In regions with many young, massive stars, a much larger scale version
of such a wind bubble may result, as multiple stellar winds and
especially supernovae contribute to the development of a single large
cavity. The evolution of such ``superbubbles'' can be largely understood
simply by extending stellar wind bubble theory, scaling upward to the
mass and energy input from a starburst (Chevalier \& Clegg 1985;
MacLow \& McCray 1988; Tomisaka \& Ikeuchi 1988). Scaled to a
canonical supernova rate $\nusn\sim 1\pyr$, the mechanical energy
and mass input rates are
\be
\Lsn \sim 3\times
10^{43}\left({\nusn\over1\;\pyr}\right)\left({\Esn\over 10^{51}\;{\rm
erg}}\right)\; \lum
\ee
\be
\mdotsn \sim 10\left({\nusn\over1\;\pyr}\right)\; \mslpyr 
\ee
For a uniform ambient density $n_o$, the radius, expansion
velocity, internal pressure, and thermal energy of the
superbubble are (\eg Heckman \etal 1996)
\be
R\sim 1.1\left({L_{43.5}\over n_o}\right)^{1/5}t_7^{3/5}\;{\rm kpc} 
\ee
\be
V=\dot R\sim 200\left({L_{43.5}\over n_o}\right)^{1/5}t_7^{-2/5}\;\kms 
\ee
\be
\tilde P={P\over k}\sim 5\times 10^6 L_{43.5}^{2/5}n_o^{3/5}
t_7^{-4/5} \;\nt
\ee
\be
E_{\rm th}\sim 4\times 10^{57}L_{43.5} t_7 \;{\rm erg .}
\ee
Finally, the soft X-ray ({\it ROSAT}-band) luminosity is approximately 
\be
L_x\sim 10^{42} L_{43.5}^{33/35}n_o^{17/35}t_7^{19/35}\;\lum\;;
\ee
this includes the effects of evaporation into the interior of the
bubble. In these equations the mechanical luminosity of the OB
association (approximated as constant) is $\Lsn=3\times
10^{43}L_{43.5}$ \lum\ and the age of the superbubble is $t=10^7t_7$
yr. 

The expansion velocity (17) is generally supersonic, so the
superbubble drives a shock into the surrounding ISM. The resulting
shock luminosity depends on both the shock velocity and the
ambient density:
\be
L_{\rm sh}\approx 2\times 10^{40} V_{100}^3 n_o A_{\rm sh}\;\lum\,
\sim 2\times 10^{42} L_{43.5} \;\lum
\ee
where the shock velocity $V_{\rm sh}=100 V_{100}\;\kms$ and the shock
area $A_{\rm sh}$ is in kpc$^2$ (Dopita \& Sutherland 1996); the
second line assumes $V_{\rm sh}$ is equal to the bubble
expansion speed and uses equations (16) and (17). For shock speeds in
the range $V_{\rm sh}\sim 10^2-10^3$ \kms, the resulting H$\alpha$
luminosity $L_{\rm H\alpha}\sim 0.01 L_{\rm sh} V_{100}^{-0.6}$. The
[O$\,$III] $\lambda 4959+\lambda 5007$ line luminosity (including the
contribution from the UV precursor) is comparable to the H$\alpha$
luminosity at $V_{100}\approx 2$ and gets relatively stronger as the
shock speed increases, with the ratio scaling approximately as $V_{\rm
sh}^{2.3}$ (Dopita \& Sutherland 1996).

The finite scale height of the gas disk means that the superbubble can
expand entirely out of the disk, provided it is energetic enough. Such
``blowout'' will occur if the dimensionless parameter
\be
{\cal D}={\Lsn\over \Lism}\sim 10^4\left({H\over 100\;{\rm
pc}}\right)^{-2}\left({\tPism\over 10^7\;{\rm cm^{-3}\;K}}\right)^{-3/2}
n_o^{1/2} \gtrapprox 100
\ee
(MacLow \& McCray 1988) where $\Lism=\Pism H^3/ t_{\rm dyn}$
is the ``luminosity'' of the ISM over the dynamical timescale $t_{\rm
dyn}$, defined as the time for the bubble to reach a radius of
comparable to the scale height $H$ of the gas layer (\cf equation
[16]). Once the superbubble shell reaches an altitude $Z\gtrapprox\;$a
few $H$, it will begin to accelerate along the density gradient and
fragment due to Rayleigh-Taylor instabilities. Provided this occurs on
a timescale less than the lifetime of the OB association powering the
supershell, the continuous injection of energy and momentum from the
association will lead to the development of a galactic wind.

There is extensive evidence for such large-scale galactic
``superwinds'' (\eg Heckman, Armus \& Miley 1990; Veilleux \etal 1994;
Heckman \etal 1996), from morphologies, kinematics, and the presence
of density/pressure profiles in reasonable agreement with the
predictions of wind models (although there is not always agreement on
the latter: see Veilleux \etal 1994). In the prototypical ULIRG Arp
220, the superbubble does not appear to have blown out, as the
morphology and kinematics suggest a confined, shocked bubble. The
starburst galaxy M82, on the other hand, appears to possess a freely
expanding wind. Suchkov \etal (1996) argue that the wind in M82 must
be mass-loaded, \ie the mass flux in the wind is enhanced by a factor
of $\sim 5$ over the SNe mass deposition rate, by entrainment or
evaporation of interstellar material. The implied mass lost through
the wind is $M_W\sim 5\times 10^7 t_7$ \msol, which is comparable to
the total molecular gas mass present in the nucleus. If this is
typical of superwinds, then the fraction of available gas which is
lost from the system is large, $\fdis\sim 0.5$.

\subsection{AGN:}
\vspace{-0.02truein}
The possibility that AGN might drive powerful nuclear winds dates back
at least to Weymann \etal (1982). More recently, this idea has been
revived by Smith (1993, 1996) to explain the kinematics of the NLR and
as a mechanism for cloud confinement. In a very provocative paper,
Bicknell \etal (1998) have proposed that the radio jets in Seyfert
galaxies are actually dominated by thermal plasma, with mechanical
luminosities of order $\Lmech\sim 0.1\Lbol$; the winds start off only
mildly relativistic. Bicknell \etal make specific application to the
excitation of the NLR in NGC 1068, and find additional support for
this model from study of regions of jet-cloud interaction within the
NLR. Interestingly, in a study of soft X-ray emission from large-scale
nuclear outflows in a sample of nearby, edge-on Seyfert galaxies,
Colbert \etal (1998) conclude that the outflows must be dominated by
thermal plasma. This is an idea which is clearly deserving of far more
study. If correct, it would replace the still unanswered question ``Why
are some objects radio-loud and others radio-quiet?'' with another:
``Why are some jets dominated by nonthermal plasma?'' 

\section{Global Energetics of the ISM}
Most ULIRGs are inferred to possess large quantities ($M\gtrapprox
10^9\;\msol$) of high-density gas ($\nh\sim 10^4\;\pcubcm$),
characterized by large velocity dispersions ($\sigma$ up to $\sim
100\;\kms$); see \eg Tacconi, this volume. This implies that the
energy dissipation rate within the ISM is very large: dimensionally,
\be
\dot E_{\rm kin}\sim\eta{M\over R}\sigma^3
\sim6\times10^{42}\eta\left({M\over 10^9\;\msol}\right) R^{-1}_{100}
\,\sigma_7^3\;\lum
\ee
where $R=100 R_{100}$ pc, $\sigma=100\sigma_7$ \kms, and the
coefficient $\eta\lessapprox 1$ 
(\eg MacLow 1998). Since the total non-rotational kinetic energy of
the ISM is 
\be
E_{\rm kin}\sim {1\over 2}M\sigma^2
\sim10^{56}\left ({M\over 10^9\;\msol}\right)\sigma_7^2\;{\rm erg,}
\ee
the energy loss timescale is 
\be
\tdiss\sim{\Ekin\over\Edotkin}\sim 5\times 10^5 {R_{100}\over
\eta\sigma_7}\;{\rm yr.}
\ee
Equivalently, since the inferred area filling factor of dense ISM in
ULIRGs is close to unity (\eg Scoville, Yun, \& Bryant 1997), the
energy loss timescale (assuming strongly dissipative collisions)
$\tdiss\sim R/\sigma\sim 10^6$ $R_{100}/\sigma_7$ yr, essentially
identical with the above estimate. Since the energy dissipation
timescale is much shorter than the probable lifetime of the ULIRG, the
random bulk motions of the ISM must be continuously powered.

\subsection{Starburst:} As noted in \S 4, the mechanical energy input
from supernovae for a $L=10^{12}L_{12}$ \lsol\ ULIRG is
\be
\Lsn\sim 3\times 10^{43}\left({\nusn\over1\;\pyr}\right)
\left({\Esn\over 10^{51}\;{\rm erg}}\right)
\;\lum
\ee
which is adequate to supply the necessary kinetic energy provided
that a substantial fraction is absorbed by the dense ISM (and that
momentum transfer to the clouds is reasonably efficient); given the
large area covering factors inferred in ULIRGs (\eg Scoville \etal
1997), this seems quite reasonable.

\subsection{AGN:} Supporting the mechanical luminosity of the ISM
requires either (1) relatively efficient conversion of radiation to
mechanical energy, or (2), mechanical (momentum) input from an AGN
wind, as discussed above, with $\Lmech\gtrapprox 0.1\% \Lbol$. 

\section{Dynamical Support by Radiation Pressure}
As we have just seen, there is enough {\it energy} available from
either a starburst or an AGN to maintain the energy loss rates
estimated for the ISM in ULIRGs. A related topic is the {\it momentum}
input rate. Scoville \etal (1995) suggested that radiation pressure
might contribute significantly to the support of the gas against
gravity in the $z=2.3$ ULIRG F10214+4724, in order to 
reconcile the CO-derived gas mass with a dynamical mass (estimated
from the CO source size and line width) which is an order of magnitude
smaller. 

Assume that a cloud at radius $r$ absorbs all of the incident flux
(and therefore photon momentum) incident upon it from the central
source of radiation. In equilibrium, the condition that the radiation
provide enough pressure to support the cloud against gravity is just
\be
{GM\mu \mh\Nh\over r^2}={L\over 4\pi r^2 c}
\ee
where $M$ is the total mass enclosed within radius $r$, $\mu$ is the
mean atomic weight in units of the hydrogen mass \mh, and \Nh\ is the
hydrogen column density through the cloud. Simple rearrangement shows
that the cloud column density which can be supported is
\be
N_{\scriptscriptstyle {\rm H}}^{\rm rad}\sim {L\over 4\pi G c\mu\mh
M} \simeq 3.5\times 10^{22}\left({L/M\over
10^3}\right)\;\psqcm
\ee
which depends on the luminosity to mass ratio, here expressed in solar
units. (Note that although I have done the calculation by balancing
the radial gravitational attraction against the radiation pressure
from a central source, the identical result is obtained by calculating
the column which can be {\it vertically} supported above a radiating
slab, which may be a more relevant geometry for starburst-powered
ULIRGs).

The limiting column density which can be supported is much smaller (by
a factor $10-100$) than the area-averaged molecular column densities
which have been derived in the nuclei of ULIRGs, for which the values
of $L/M$ generally lie in the range $10^2-10^3$, suggesting that
radiation pressure support is not important. The expression for the
critical column (28) assumes that the photons give up all of their
momentum immediately, in other words, it has been assumed that the ISM
is optically thin to the re-radiated photons (generally in the mid- to
far-infrared). If the ISM is optically thick to these photons, then
multiple scattering of the radiation will occur. However, this will
{\it not} increase the column which can be supported. Assume that the
medium is very optically thick, so that a large number of scatterings
occur. In this case the radiation field will become nearly
isotropic. The degree of anisotropy (which determines how much work
the radiation field can do against gravity) is then set by the net
flux outward; this is (in spherical geometry) precisely $L/4\pi R^2$,
and so the value of $N_{\scriptscriptstyle {\rm H}}^{\rm rad}$ is
unaltered.

\section{Pressure in the ISM}
In any model for ULIRGs, an inevitable result is that the pressure in
the ISM must be extremely high:
\begin{itemize}

\item[$\bullet$] Simply from the derived gas densities and observed
turbulent velocities, the inferred turbulent pressures must be high:
\be
\tPturb\sim 5\times 10^8\left({n\over
10^4\;\pcubcm}\right)\left({\sigma_{\rm turb}\over 30\;\kms}\right)^2
\;\nt .
\ee

\item[$\bullet$] If the nuclear ($R\sim 100$ pc) ISM essentially
radiates like a blackbody, as proposed by Downes \& Solomon (1998),
then the radiation pressure in this region
\be
P_{\rm rad}={u\over 3}\sim {\Lbol\over 12\pi R^2 c}
\ee
(where $u$ is the radiation energy density) and so therefore
\be
\tilde P_{\rm rad}\sim 3\times 10^8 L_{12}
R_{100}^{-2}\;\nt\;.
\ee

\item[$\bullet$] In an AGN-powered ULIRG, the pressure of radiation from
the central source of luminosity will force the gas pressure to large
values: 
\be
\tilde P_{\rm rad}\sim 8\times 10^7\left({f_{\rm abs}\over 0.1}\right)
L_{12} R_{100}^{-2}\;\nt
\ee
where $f_{\rm abs}$ is the fraction of the incident radiation which is
absorbed in a thickness $\delta r\ll r_c$, the cloud radius
(see Barvainis \etal 1997).

\item[$\bullet$] In a starburst-powered ULIRG, with enough energy input
from SNe to drive a superwind, the pressure in the injection region
(\ie on the spatial scale of the starburst) is
\be
P_o\sim 0.1\dot M^{1/2}L_{\scriptscriptstyle {\rm SN}}^{1/2} 
R_{\scriptscriptstyle {\rm SB}}^{-2}
\ee
(Chevalier \& Clegg 1985) where $R_{\scriptscriptstyle {\rm SB}}$ is
the radius of the starburst region; hence
\be
\tilde P_o\sim 3\times 10^8 \left({\mdotsn\over \mslpyr}\right)
\left({\Lsn\over 10^{43.5}\;\lum}\right)\left({R_{\scriptscriptstyle
{\rm SB}}\over 100\;{\rm pc}}\right)^{-2}\;.
\ee
\end{itemize} 

All of these estimates agree to order of magnitude, implying that the
pressure in the ISM $\tPism\sim 10^8\;\nt$ in ULIRGs. This is $3-4$
orders of magnitude larger than the pressure in the average ISM in the
solar neighborhood; only the densest cloud cores approach such high
pressures. Similarly high pressures have been inferred in more modest
starburst galaxies such as NGC 1808 and NGC 3256 (\eg Aalto \etal
1994, 1995).

\section{Local Energetics: Energy Input to Molecular Clouds}
\subsection{Starburst:}
The power input from an OB association in the form of radiation,
stellar winds, and supernovae is
\be
\Edotob\sim 6\times 10^{35} N_*\;\lum
\ee
(McCray \& Kafatos 1987) where $N_*$ is the number of stars in the
association with $M> 8\;\msol$. For lower and upper stellar mass
cutoffs of $m_l=1$ and $m_u=60$ \msol, respectively, (which gives 1
SNe for every $\sim 100\;\msol$ turned into stars) the energy input
to the cloud from which the association formed over the cloud's lifetime
$\Delta t$ is
\be
\Eob\sim 5\times 10^{51}\left({\epsilon_*\over 0.01}\right)
\left({M_c\over 10^5\;\msol}\right)\left({\Delta t\over 10^7\;{\rm
yr}}\right)\;{\rm erg}
\ee
where $M_c$ is the cloud mass and $\epsilon_*$ is the star formation
efficiency as before. 

The cloud binding energy (for mean cloud density $\nh=10^4 n_4\;\pcubcm$)
\be
E_{\scriptscriptstyle {\rm B}}\sim {GM_c^2\over r_c}
\sim 2\times10^{50}\left({M_c\over 10^5\;\msol}\right)^{5/3}
n_4^{1/3}\;{\rm erg}\;;
\ee
the ratio of energy input to binding energy is thus
\be
{\Eob\over\Eb}\sim 25 \left({\epsilon_*\over 0.01}\right)
\left({M_c\over 10^5\;\msol}\right)^{-2/3}
n_4^{-1/3}\left({\Delta t\over 10^7\;{\rm
yr}}\right)
\ee
which depends relatively weakly on the assumed cloud parameters. Thus,
in principle, only a few percent of the available power needs to be
input to the cloud in order to disperse it (see \eg Williams \& McKee
1997). However, most of this energy will of course be radiated away.

\subsection{AGN:}
If the clouds in the ISM have a direct view of the active nucleus,
then the integrated energy input (assuming $\Delta t\le$
the lifetime of the AGN) is simply
\bea
E_{\scriptscriptstyle {\rm AGN}}&\sim& {\Lbol\over 4\pi R^2} A_{\rm cl}
\Delta t \nonumber \\
&\sim& 10^{57} L_{12} R_{100}^{-2}
f_{\rm abs}\left({M_c\over
10^5\;\msol}\right)^{2/3} n_4^{-2/3}
\left({\Delta t\over 10^7\;{\rm yr}}\right)\;{\rm erg}
\eea
where $A_{\rm cl}$ is the projected area of the cloud.
Thus the ratio of AGN energy input to cloud binding energy is
\be
{\Eagn\over\Eb}\sim 10^6 L_{12} R_{100}^{-2}
f_{\rm abs}\left({M_c\over
10^5\;\msol}\right)^{-1} n_4^{-1}\left({\Delta t\over 10^7\;{\rm
yr}}\right) 
\ee
which is several orders of magnitude larger than the energy input
calculated above for a starburst. (Since the bolometric luminosities
are the same in both cases, the same amount of radiation -- mostly
re-radiated in the far-IR -- must of course be incident on a typical
cloud in a starburst-powered ULIRG. Precisely because this radiation
is dominantly at infrared wavelengths, however, it will have far less
impact on the ISM in a ULIRG than the radiation from an AGN, serving
mostly to increase the typical ISM temperatures.)
Evaporation and ablation of clouds which are exposed to the AGN can be
very significant, leading to destruction of the molecular ISM
(Begelman 1985). 

Because of the large gas column densities inferred in the nuclei of
ULIRGs, however, it is likely that most of the ISM does not have an
unobscured view of the AGN. Here the fact that the spectrum of
an AGN extends to high energies ($E\sim 100$ keV) is of great
importance, as these photons can penetrate large column densities
before being absorbed. This means that the amount of energy deposited
by X-rays can be very large even if there is a significant column
between a cloud and the radiation source: the energy deposition rate
per unit volume is approximately
\be
nH_x\sim 3\times 10^{-17}\left({f_x\over 0.1}\right)L_{12}
R_{100}^{-2}\left({\Natt\over 10^{22}\;\psqcm}\right)^{-1}
n_4\;\einpuv
\ee
(Maloney, Hollenbach, \& Tielens 1996, hereafter MHT), where $f_x$ is
the fraction of the bolometric luminosity emitted in the $1-100$ keV
energy band, and \Natt\ is the column density of neutral hydrogen
between the cloud and the AGN. For shielding columns $\Natt\gtrapprox
10^{22}$, only hard X-ray ($E > 1$ keV) photons will be
present. Unlike the case of the photodissociation regions (PDRs)
powered by stellar UV photons, where dust absorption leads to
exponential attenuation of the heating and ionization rates with
depth, the XDRs powered by AGN are characterized by an only linear
decrease of the energy input with increasing column density (the exact
scaling with column density depends weakly on the spectral index of
the incident radiation field: MHT). The total energy input to the
cloud is then
\be
\Exagn\sim 10^{55}\left({f_x\over 0.1}\right)L_{12}
R_{100}^{-2}N_{22}^{-1}\left({M_c\over 10^5\;\msol}\right)
\left({\Delta t\over 10^7\;{\rm yr}}\right)\;{\rm erg}
\ee
where the shielding column has been scaled to $\Natt=10^{22}N_{22}$
\psqcm. Compared to the cloud binding energy, 
\be
{\Exagn\over\Eb}\sim 10^5 \left({f_x\over 0.1}\right)L_{12}
R_{100}^{-2}N_{22}^{-1}\left({10^5\;\msol\over M_c}\right)^{2/3}
n_4^{-1/3}\left({\Delta t\over 10^7\;{\rm yr}}\right)\;{\rm erg .}
\ee
Thus, even if the shielding colummn between the X-ray source and a
cloud is $\Natt\sim 10^{24}$ \psqcm, the X-rays from the AGN can still
dump $\sim 10^3$ times the cloud binding energy into a cloud over its
lifetime. As in the case of a starburst-powered ULIRG, most of this
energy is simply radiated away (although photoevaporation will occur
if the clouds are heated enough to raise the sound speed above the
escape speed). It does mean, however, that it is extremely easy to
``light up'' the interstellar medium in a ULIRG, simply in consequence
of the energy input rate, whether it is powered by an AGN or a
starburst.

\section{X-Ray Irradiation in AGN-Powered ULIRGs}
As alluded to above, the large column densities which hard X-ray
photons can traverse before being absorbed, combined with the large
energy per photon, allow X-ray luminous AGN to have a profound impact
on the physical and chemical state of the ISM in AGN-powered
ULIRGs. In this section I sketch briefly some of the results and their
observational consequences.

\subsection{Physical and Chemical Structure}
Detailed models of the physics and chemistry of XDRs have been
discussed by MHT. The important parameter is $H_x$, the X-ray energy
deposition rate per hydrogen nucleus. The heating rate (as noted
earlier) and the molecule destruction rates scale as $nH_x$, whereas
the cooling and molecule formation rates are generally $\propto n^2$;
thus the physical and chemical state scales approximately with
$H_x/n$. It is convenient to express this in terms of an X-ray {\it
ionization parameter} (the ratio of photon flux to gas density), 
\be
\xi={L_x\over nR^2}={4\pi F_x\over n}
\ee
where $F_x$ is the X-ray flux incident on the face of the cloud. For
optically thin gas, the ionization parameter determines the
physical state (\eg Tarter, Tucker \& Salpeter 1969). In the case of
interest here, where the clouds are optically thick to the
X-rays (up to some maximum energy), the physical conditions depend on
an effective ionization parameter,
\be
\xi_{\rm eff}={4\pi F_x\over n N_{\rm att}^\alpha} 
\simeq 0.1 {L_{44}\over n_4 R_{100}^2 N_{22}^\alpha}
\ee
where the $1-100$ keV X-ray luminosity is $L_x=10^{44}L_{44}$ \lum\
and, as noted earlier, the scaling index $\alpha\sim 1$ for typical
AGN spectral indices. This effective ionization parameter is directly
proportional to $H_x/n$, with $H_x/n\sim 4\times 10^{-25}\xieff$ erg
cm$^3$ s$^{-1}$.

\begin{figure}
\plotfiddle{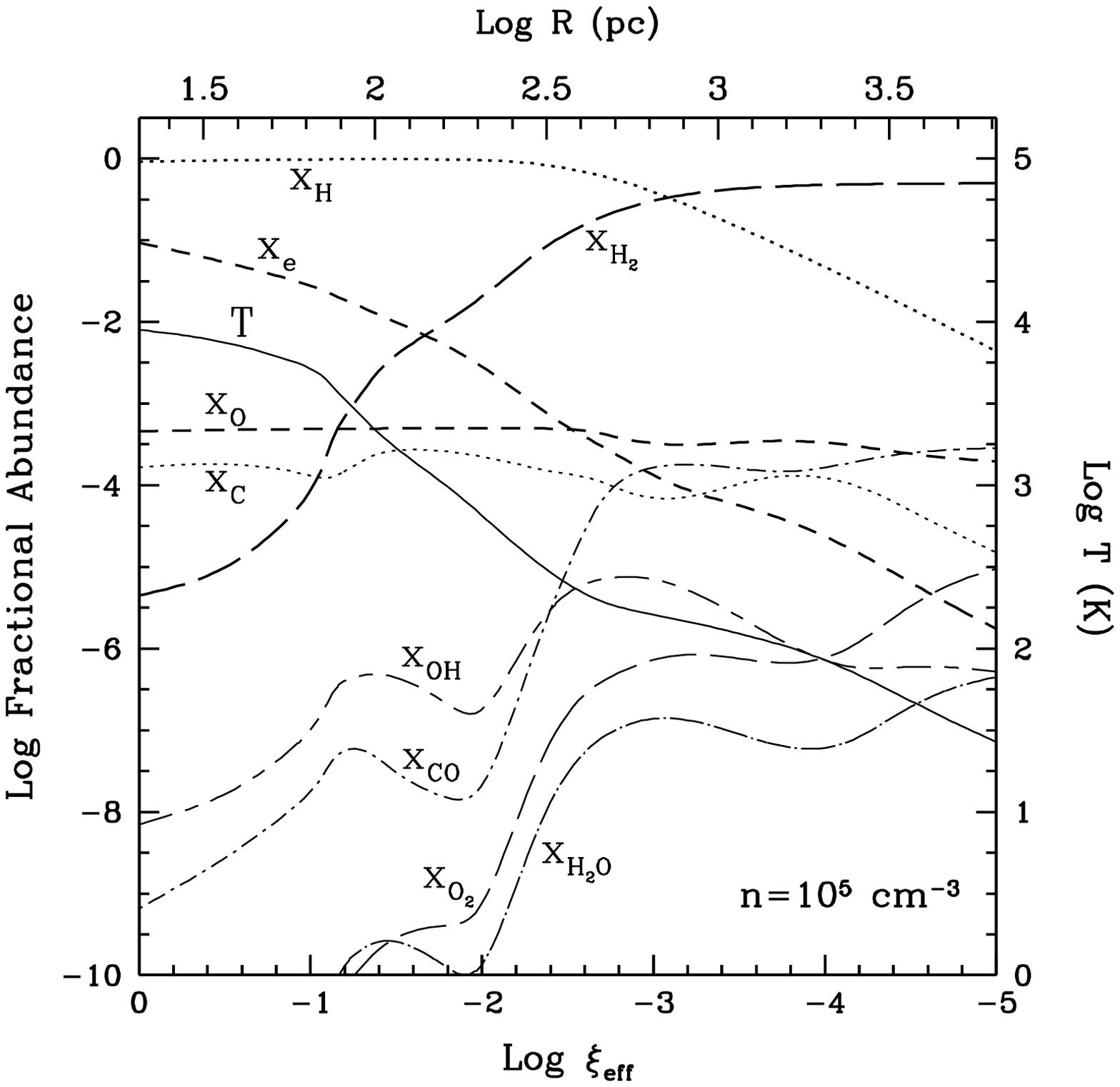}{4.0in}{0}{73}{73}{-240}{-180}
\caption{XDR physical conditions as a function of radius (or \xieff),
for $L_x=4\times 10^{44}$ erg s$^{-1}$, $N_{\rm att}=10^{22}\psqcm$, and
$n=10^5\pcubcm$.} 
\end{figure}

Figure 1 shows the physical and chemical structure of an XDR
 as a function of \xieff\ or, equivalently, distance from the
X-ray source. This calculation assumes $L_x=4\times 10^{44}$
\lum, or $f_x\approx 0.1$ for $\Lbol=10^{12}$ \lsol.
The gas density is constant at $n=10^5$ \pcubcm, and the column
between the X-ray source and the gas is $\Natt=10^{22}$ \psqcm. (Cloud
column densities $N_{\rm cl}=2\times 10^{22}$ \psqcm\ and linewidths
$\Delta V=5$ \kms\ have been assumed in calculating the cooling of
optically thick species.) Close to the X-ray source, the gas is warm,
$T\sim 10^4$ K, atomic, and weakly ionized, with ionization fraction
$x_e\sim 0.01-0.1$. With increasing distance (or increasing column to
the X-ray source) the temperature and ionization fraction decreases
and the abundances of molecular species increase; at the lowest values
of \xieff, the temperature has dropped to $T\approx 25$ K, and nearly
all of the gas-phase carbon is in CO. Atomic oxygen is abundant over
the entire range of \xieff.

\subsection{Molecular Survival}
Figure 1 shows that, for a gas density $n=10^5\;\pcubcm$, CO only
becomes an abundant species for $R\gtrapprox 100$ pc, whereas many
ULIRGs show bright CO emission on smaller spatial scales. What
densities are needed for gas to be molecular in an AGN-powered ULIRG? 

Define the fractional abundance of a species by $x_i=n_i/n$, where
$n_i$ is the number density of species $i$ and $n$ is the total
density of hydrogen nuclei, $n=\nh+2\nhtwo$; note that this definition
means that the molecular hydrogen fraction $\xhtwo=0.5$ if all of the
hydrogen is in molecular form.
In table 1, I give the density $n_m$ required in order for the CO
abundance $\xco=10^{-4}$ for three different
distances, $R=10$, 100, and 1000 pc from the nucleus, and three
different values of the shielding column, $\Natt=10^{22}$, $10^{23}$,
and $10^{24}$ \psqcm. Also listed in the table are the corresponding
molecular hydrogen fraction \xhtwo\ and the gas temperature $T$. 

\vspace{-0.3truein}
\begin{table}[htb]
\begin{center}
\caption{Densities for CO Survival}
\begin{tabular}{cccc}
\hline
\Natt&$n_m$&\xhtwo&T(K)\\
\hline
\multicolumn{4}{c}{}\\
&\multicolumn{2}{c}{$R=10$ pc}&\\
$\qq 10^{22}\qq$&$\qq 1.5\times 10^8\qq$&$\qq 0.07\qq$&$\qq903\qq$\\
$10^{23}$&$2.5\times 10^7$&$0.10$&694\\
$10^{24}$&$4.8\times 10^6$&$0.13$&545\\
\multicolumn{4}{c}{}\\
&\multicolumn{2}{c}{$R=100$ pc}&\\
$\qq 10^{22}\qq$&$\qq 1.9\times 10^6\qq$&$\qq 0.14\qq$&$\qq409\qq$\\
$10^{23}$&$3.3\times 10^5$&$0.19$&233\\
$10^{24}$&$8.5\times 10^4$&$0.27$&146\\
\multicolumn{4}{c}{}\\
&\multicolumn{2}{c}{$R=1000$ pc}&\\
$\qq 10^{22}\qq$&$\qq 3.7\times 10^4\qq$&$\qq 0.30\qq$&$\qq125\qq$\\
$10^{23}$&$4.5\times 10^4$&$0.47$&74\\
$10^{24}$&$1.4\times 10^4$&$0.48$&42\\
\hline
\end{tabular}
\end{center}
\end{table}

\vspace{-0.1truein}
Table 1 clearly demonstrates how powerful the effect of an X-ray
luminous AGN on the ISM can be. At a distance of 100 pc from the
nucleus, even with a column of $\Natt=10^{24}$ \psqcm\ in front of the
X-ray source, the gas density must be $n\gtrapprox\;$a few$\,\times
10^4$ \pcubcm\ in order for CO to be abundant: for a density $n=2\times
10^4$, the CO abundance $\xco < 10^{-6}$. The temperatures are 
considerably higher than the typical ISM in the Milky Way; note also
that the thermal balance has been calculated including only the
heating due to X-rays. Clearly, high gas densities are mandatory for
molecular survival in AGN-powered ULIRGs, even when the AGN is buried
beneath a large column density of gas and dust.

\section{Far-Infrared Fine-Structure Lines}
In the post-{\it ISO} era, surely no one needs to be reminded of the
wealth of information provided by far-infrared fine-structure line
observations, a point that was made repeatedly throughout this
meeting. As discussed by MHT, XDRs are copious sources of emission in
these lines.

Figure 2 shows the emergent surface brightnesses in a number of
important mid- and far-infrared lines, for the same XDR model
parameters as used for Figure 1. Also plotted is the total
far-infrared continuum surface brightness from grain emission. Of
particular note are: (1) the extremely high line surface brightnesses
which can be produced; (2) major cooling lines such as [O$\;$I]
63$\mu$m and [Si$\;$II] 35$\mu$m are bright over nearly the entire
parameter space; and (3) compared with the far-IR continuum produced
in the XDR, the surface brightnesses in these same lines can be
extraordinarily high, with line-to-continuum ratios approaching
$10\%$. In contrast, the line-to-continuum ratios produced in PDRs are
usually of order $0.1\%$, and rarely exceed $1\%$ (\eg Tielens \&
Hollenbach 1985). This difference is a reflection of the fact that in
PDRs the gas heating is limited by the grain photoelectric heating
efficiency; in contrast, in XDRs about as much energy is deposited
into the gas as into the grains, so that major cooling transitions can
carry a large fraction of the total deposited energy (see MHT for
details). The other important point to note is that all of these lines
arise from neutral or singly-ionized species; because of the large
column densities which X-ray photons can traverse, the volume of the
XDR can be vastly larger than the volume of any high-ionization region
produced by the AGN.

\begin{figure}
\plotfiddle{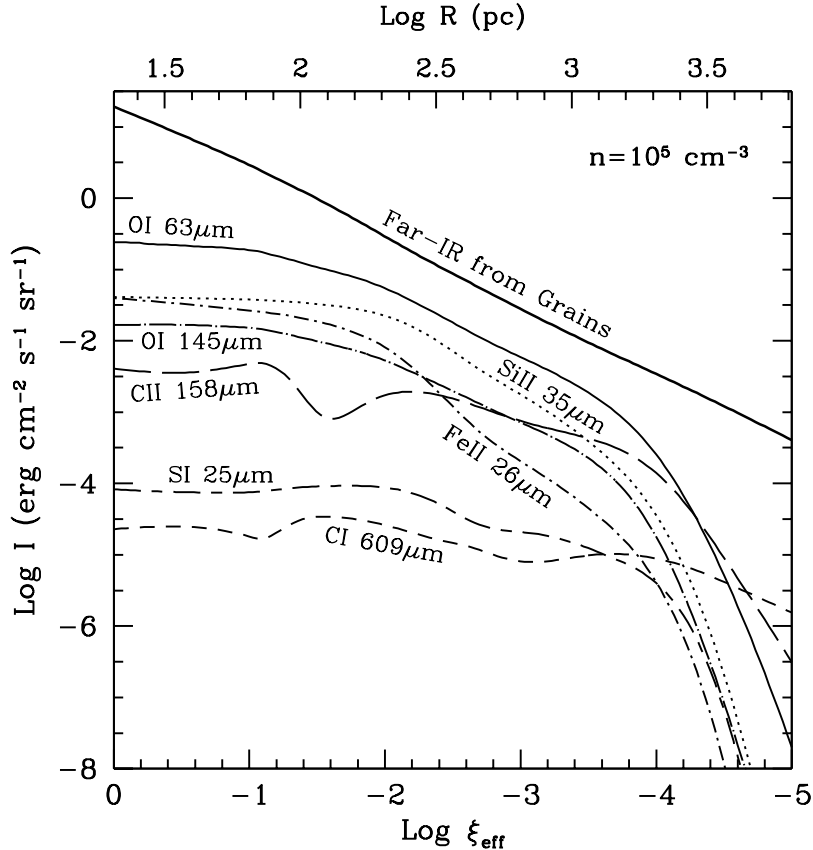}{4.0in}{0}{73}{73}{-240}{-180}
\caption{Emergent surface brightnesses in important mid- and
far-infrared lines for the XDR model of Figure 1.} 
\end{figure}

\subsection{Particularly Interesting Lines}
A number of other infrared transitions are of particular interest for
ULIRGs. The near-infrared vibration-rotation lines of molecular
hydrogen have been detected from a number of ULIRGs (most
spectacularly in the not-quite-an-ULIRG NGC 6240); the $1.64\mu$m
fine-structure line of [Fe$\:$II] is frequently detected as
well. Observations with {\it ISO} have produced detections of the
[Ne$\:$II] $12.8\mu$m line in a number of ULIRGs as well, and
the first extragalactic detections of the pure rotational
transitions of H$_2$, including observations of the archetypical ULIRG
Arp 220 (Sturm \etal 1996).

In Figure 3 I have plotted the emergent surface brightnesses for the
$v=1-0$ and $v=2-1$ S(1) lines of molecular hydrogen, the $1.64\mu$m
[Fe$\:$II] line, Br$\gamma$, and the [Ne$\:$II] $12.8\mu$m line. As in
Figure 2, the high line surface brightnesses are striking. The extent
of the [Fe$\;$II] emission is limited by gas temperature, since the
upper state lies $E/k\sim 10^4$ K above the ground state. The drop-off
in molecular hydrogen emission at $\log\xieff\approx 2$ is also due to
declining $T$ (the $v=1$ and $v=2$ levels lie approximately 6000 and
12,000 K above ground, respectively); the secondary rise is due to
direct nonthermal excitations of the levels. For a fixed density, the
[Fe$\:$II] emission always lies interior to the molecular hydrogen
emission (see also Maloney 1997). The run of density with $\xieff$ (or
radius) will determine the ratios of the H$_2$ and [Fe$\:$II] lines
with respect to each other and to Br$\gamma$. 

\begin{figure}
\plotfiddle{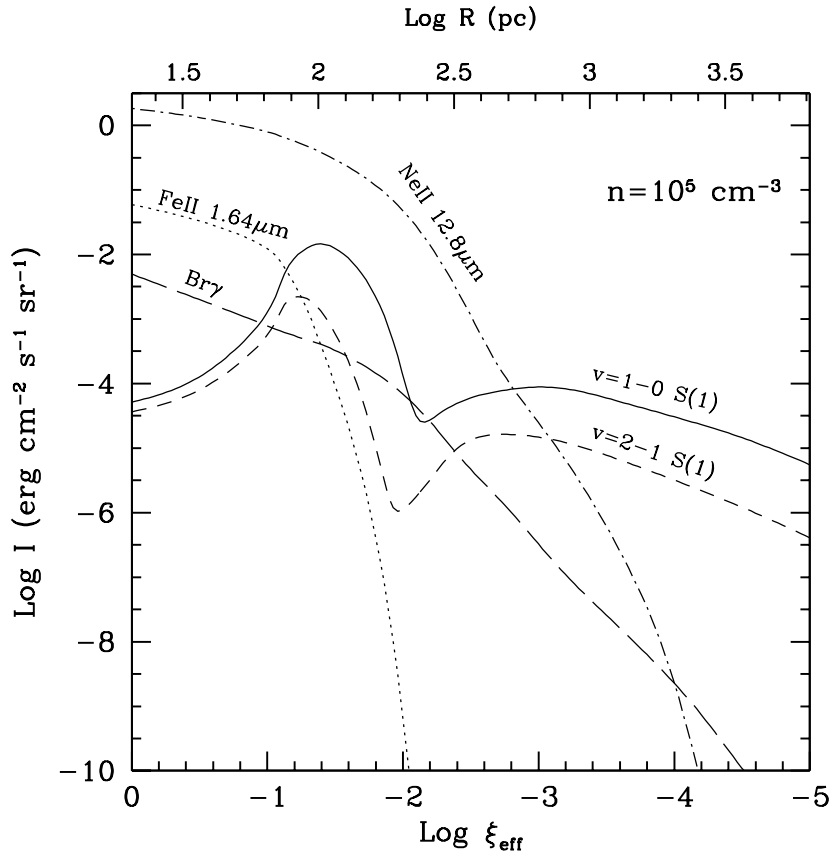}{4.0in}{0}{73}{73}{-240}{-180}
\caption{Emergent surface brightnesses in several important
near- and mid-infrared lines for the XDR model of Figure 1.} 
\end{figure}

An additional interesting result is the intensity of the [Ne$\:$II]
$12.8\mu$m line. This is the single most important coolant over a
substantial fraction of the plotted parameter space; as much as $\sim
15\%$ of the deposited X-ray energy emerges in this line alone. The
resulting surface brightnesses are extraordinarily high, and the total
line luminosity can be a substantial fraction of the total hard X-ray
luminosity. 

The pure rotation lines of molecular hydrogen are of particular
interest because they are much less subject to extinction than the
near-infrared lines, and because they provide a tracer of warm (few
hundred K) molecular gas, unlike the vibration-rotation bands which
require $T\gtrapprox 2000$ K for effective excitation. The S(0), S(1),
S(3), and S(5) lines have been detected in NGC 1068 (Lutz \etal 1997)
and the S(1) and S(5) lines (at 17 and $6.9\mu$m, respectively) have
been detected in Arp 220 (Sturm \etal 1996). 

In Figure 4 I have plotted the surface brightnesses of the S(0), S(1),
S(3), and S(5) lines for the XDR model of Figure 1. Current
observations lack the spatial resolution to do more than measure the
total line luminosities. However, even this should prove extremely
useful for constraining models of the ISM in ULIRGs, and XDR models in
particular (Lutz \etal 1997).

\begin{figure}
\plotfiddle{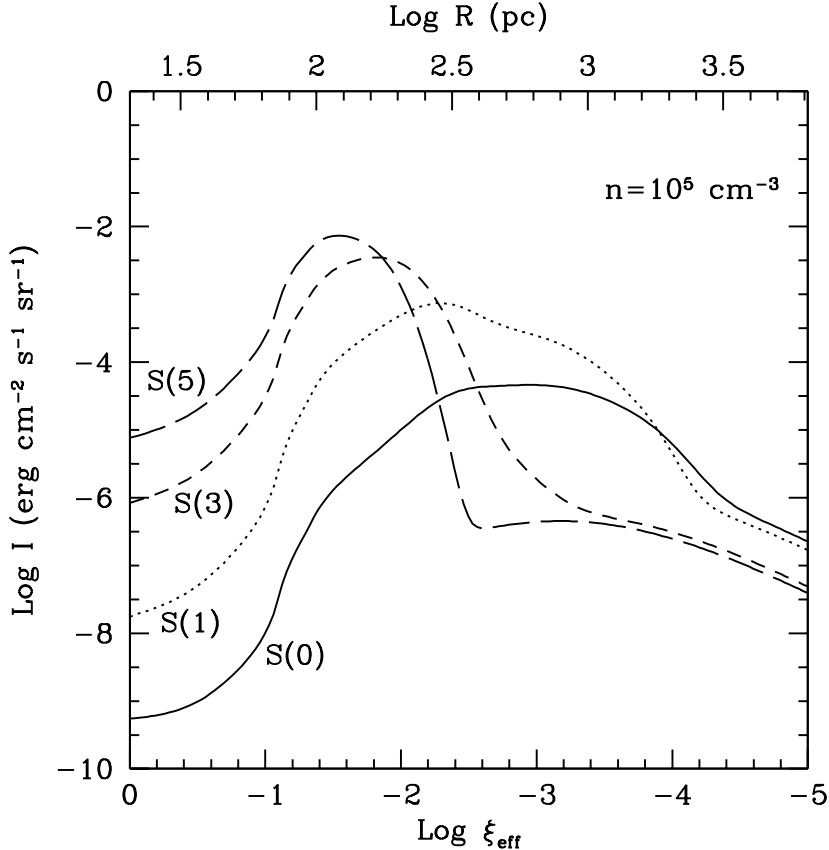}{4.0in}{0}{73}{73}{-240}{-180}
\caption{Emergent surface brightnesses in the S(0), S(1), S(3) and
S(5) pure rotation lines of H$_2$, for the XDR model of Figure 1.} 
\end{figure}

\subsection{The HCN/CO Ratio}
In ULIRGs, the ratio of the HCN \jonezero\ line to that of CO is
typically $\sim 0.2$ (Solomon, Downes, \& Radford 1992), an order of
magnitude larger than the value typically observed in normal
spirals. In the nuclear ($R\lessapprox 100$ pc) region of NGC 1068,
this ratio reaches unity (Tacconi \etal 1994). Such high values have
been interpreted as meaning that the density is high ($n\gtrapprox
10^5$ \pcubcm), so that both the HCN and CO lines are approximately
thermalized, and that both of the transitions are optically thick, or
nearly so. However, the inferred high densities must be regarded with
some caution, as the HCN rotational transitions can be pumped through
a bending mode via absorption at $14\mu$m; this process is plausibly
important in both starbursts and AGN (Aalto \etal 1995; Barvainis
\etal 1997).

The line center optical depth for a transition at frequency $\nu$ is
\be
\tau_o={c^3\over 8\pi\nu^3}A_{ji}{g_j\over g_i}{N_i\over \Delta V}
\left(1-e^{-h\nu/kT_{\rm ex}}\right)
\ee
where $A_{ji}$ is the Einstein $A-$coefficient, $j$ and $i$ denote the
upper and lower states, the $g$s are the statistical weights, $N_i$ is
the column density in the lower state, $\Delta V$ is the linewidth,
and $T_{\rm ex}$ is the excitation temperature characterizing the
relative populations of levels $i$ and $j$. Assuming equal excitation
temperatures for the two transitions and that $h\nu/kT_{\rm ex}\ll 1$,
the ratio of $\tau_o=1$ columns for the \jonezero\ transitions is then
\be
{N_o({\rm HCN})\over N_o({\rm CO})}={|\muco|^2\nuco\over
|\muhcn|^2\nuhcn} \simeq 1.8\times10^{-3}.
\ee
This ratio is much less than unity because the $A-$coefficient for the
HCN transition is so much larger than for the corresponding transition
of CO, in consequence of its much larger dipole moment $\mu$. Thus, if the
HCN abundance is more than a fraction of a percent of the CO
abundance, the HCN \jonezero\ transition will be optically thick if
the CO transition is. Of course, if the CO \jonezero\ optical depth is
very large, then a much smaller HCN abundance will still lead to an
optically thick HCN line. Furthermore, it is very important to note
that equation (47) gives the ratio of required column densities in
$J=0$. Because of its much lower critical density, it is likely that
the partition function for CO will be close to the LTE limit for the
relevant conditions, while that of HCN will not, so that the CO
molecules will be distributed over a much larger number of rotational
states. Hence the ratio $N_o/N_{\rm tot}$ will be much larger for HCN
than for CO, also reducing the required HCN abundance.

In the quiescent interstellar medium in the solar neighborhood, the
optical thickness condition is not fulfilled, as the HCN abundance is
usually $\xhcn\sim 10^{-4}\xco$. Furthermore, there are indications
that in very active galactic nuclei (either high SFR or AGN), the
optical depths in the CO lines are only modest (Aalto \etal 1995;
Barvainis \etal 1997), so that the HCN lines are not optically thick
in reflection of very optically thick CO lines. However, the increased
ionization rate in an XDR can lead to an enhanced abundance of HCN
relative to CO (Lepp \& Dalgarno 1996). In Figure 5 I show the
abundances of CO, HCN, and the ratio $\xhcn/\xco$ as a function of
\xieff\ and distance. The AGN parameters are the same as previously,
but I have increased the density to $n=10^8$ \pcubcm\ simply to make
the gas dominantly molecular within 100 pc of the nucleus. The HCN/CO
ratio is greater than $10^{-3}$ over about a factor of three in
radius. Thus in an XDR the HCN abundance can be elevated sufficiently
relative to CO to explain the observed line ratios. (In hot
star-forming cores, the HCN/CO ratio also reaches values of order
$10^{-3}$ [Blake \etal 1987], so elevated values may be expected in
starburst-powered ULIRGs as well.)

\begin{figure}
\plotfiddle{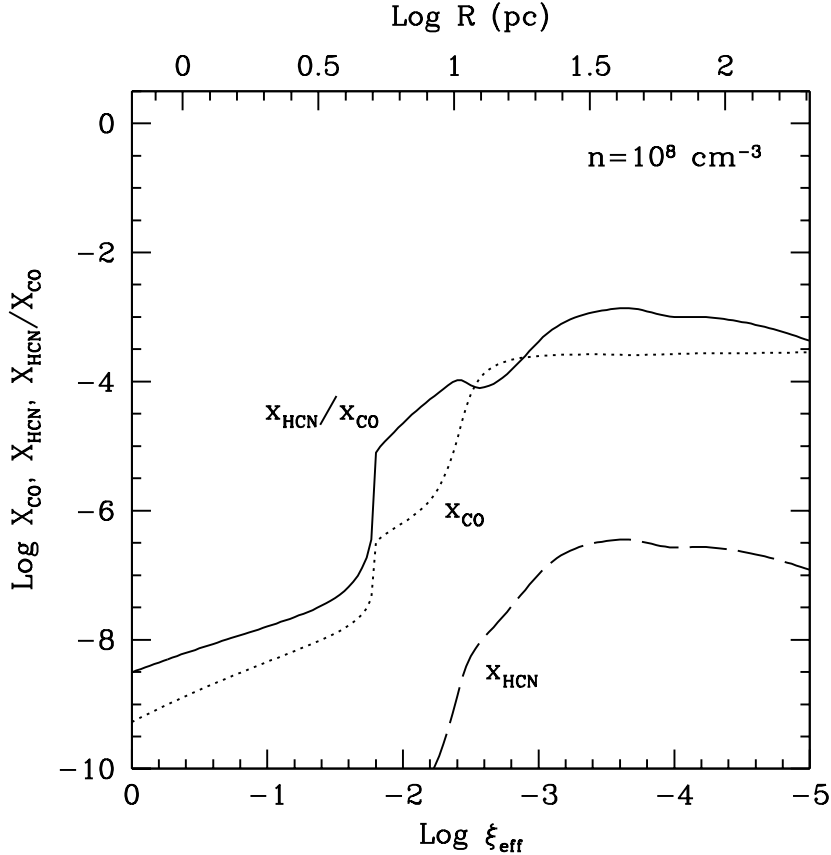}{4.0in}{0}{73}{73}{-240}{-180}
\caption{HCN and CO abundances and their ratio as a function of
\xieff. The assumed AGN and cloud parameters are the same as Figure 1,
except that the assumed density has been increased to $n=10^8$
\pcubcm.} 
\end{figure}

\section{PAH Emission}
The mid-IR emission features attributed to polycyclic aromatic
hydrocarbons (PAHs) have been used to discriminate between AGN and
starbursts in ULIRGs (\eg Lutz \etal 1998), based on the observation
that these features are weak or absent in classical AGN but generally
strong in starbursts. Does the presence of PAH features imply the
absence of an AGN?

Voit (1992) has investigated the effects of X-ray irradiation on
PAHs. Absorption of an energetic photon leads to destruction of the
PAH, through two processes: (1) photo-thermal dissociation (\ie
evaporation), and (2) Coulomb ``explosion'', in which the PAH is
doubly ionized by the incident photon and subsequently fragments as
the repulsive Coulomb force effectively reduces the binding energy of
the PAH. To estimate the magnitude of the destruction timescale,
assume that every absorption of an X-ray photon by a PAH leads to its
destruction\footnote{This is likely to overestimate the PAH
destruction rate; far more laboratory work (\eg Jochims \etal 1996)
needs to be done in this area.}:
\be
\tau_{\rm xd}\sim 2 \left({f_x\over 0.1}\right) L_{12} R_{100}^{-2}
N_{22}^{1.6}\;{\rm yr}
\ee
assuming a flat ($\nu L_\nu={\rm constant}$) spectrum. Even if
$\Natt=10^{24}$ \psqcm, the PAH X-ray destruction time is only
$\tau_{\rm xd}\sim 3000$ years at 100 pc.

To determine whether PAHs can survive in the face of such short
destruction timescales, we need to compare $\tau_{\rm xd}$ with the
rate at which the PAHs re-accrete material (this is dominated by
accretion of C$^+$ if the PAHs are neutral). Since a photo-destruction
rate is balanced against a density-dependent reaction rate, this
again leads to a critical ionization parameter at which $\tau_{\rm
xd}=\tau_{\rm acc}$, which, using the results of Voit (1992), is
\be
U_{\rm cr}\equiv {n_{\rm ph}\over \nh}\sim 5\times 10^{-9}
\left({Y\over 0.3}\right) N_{50}^{1/2} f_{\rm PAH^o}{1\over
\Sigma(\Nh)}
\ee
where $Y$ is an effective sticking coefficient for accretion, the
number of carbon atoms in the PAH is $\Nc=50
N_{50}$, and \fpah0\ is the fraction of PAHs which are neutral.
$\Sigma(\Nh)$ is a dimensionless, spectrum-weighted cross section
for photodissociation of carbon from a PAH, (\ie the photon
flux-weighted photodissociation cross-section, normalized to the PAH
geometric cross-section), approximately given by
$\Sigma(\Nh)\sim 4\times 10^{-5}/N_{22}$ (see Figure 3 in Voit
1992). This gives the critical ionization parameter for PAH survival
as
\be
U_{\rm cr}\sim 1.3\times10^{-4}\left({Y\over 0.3}\right) N_{50}^{1/2}
f_{\rm PAH^o} f_{\scriptscriptstyle C^+} N_{22}
\ee
where $\fcp=\xcp/\xc({\rm total})$. From this expression, we
find that PAHs will survive against X-ray destruction at a radius
(with $\fcp=f_{\rm PAH^o}=1$)
\be
R_{\rm cr}\sim 100 \left[\left({f_x\over 0.1}\right)
L_{12}\left({T\over 100\;{\rm K}}\right)\right]^{1/2}
\left({\tPism\over 10^7\;\nt}\right)^{-1/2} N_{22}^{-1/2}\;{\rm pc}.
\ee
Note the importance of large ($N_{22}\gtrapprox 1$) shielding columns
and high pressures for the survival of PAHs. Since, as we have seen,
both large column densities and high pressures are characteristic of
the ISM in ULIRGs, it is quite probable that PAHs will in general
resist destruction by X-ray irradiation even in AGN-powered ULIRGs,
and so the presence of PAH features should not be taken as
conclusive proof of the absence of an AGN.

{\it Acknowledgements} I am very grateful to John Black, David
Hollenbach, and the referee, Frank Bertoldi, for their comments, and
to NASA grant NAG5-4061 and the conference organizers for support.

\end{document}